 \documentclass[preprint2]{aastex}

\slugcomment{}

\shorttitle{Gravitationally-lensed $z\sim2$ compact quiescent galaxies}
\shortauthors{Geier et al.}

\begin{document}


\title{VLT/X-Shooter Near-Infrared Spectroscopy and HST Imaging of Gravitationally-Lensed 
$z\sim2$ Compact Quiescent Galaxies\thanks{Based on observations collected at the European 
Organization for Astronomical Research in the Southern Hemisphere, Chile, 
under programs 087.B-0812 (PI: Toft) and 073.A-0537 (PI: Kneib)}}


\author{S. Geier\altaffilmark{2,3}, 
J. Richard\altaffilmark{4}, 
A.~W.~S. Man\altaffilmark{2},
T. Kr\"uhler\altaffilmark{2},
S. Toft\altaffilmark{2},
D. Marchesini\altaffilmark{5},  
J.~P.~U. Fynbo\altaffilmark{2}
}

\affil{$^{2}$Dark Cosmology Centre, University of Copenhagen, Juliane Maries Vej 30, 2100~Copenhagen, Denmark\\
\affil{$^{3}$Nordic Optical Telescope, Apartado 474, 38700~Santa Cruz de La Palma, Spain}
\affil{$^{4}$Centre de Recherche Astronomique de Lyon, Universit\'e Lyon 1, 9 Avenue Charles Andre, 69230 Saint Genis Laval, France}
\affil{$^{5}$Department of Physics and Astronomy, Tufts University, Medford, MA 06520, USA}
\email{sgeier@astro.ku.dk}
}




\begin{abstract}
Quiescent massive galaxies at $z\sim2$ are thought to be the progenitors of present-day massive ellipticals. Observations revealed them to be extraordinarily compact. The determination of
stellar ages, star formation rates and dust properties via spectroscopic measurements has up to now only been feasible for the most luminous and massive specimens ($\sim3\times M\star$).
 Here we present a spectroscopic study of two near-infrared selected galaxies
which are close to the characteristic stellar mass $M\star$ ($\sim0.9\times~M\star$
and $\sim1.3\times~M\star$) and whose observed brightness has been boosted by
the gravitational lensing effect. We measure the redshifts of the two galaxies
to be $z=1.71\pm0.02$ and $z=2.15\pm0.01$. By fitting stellar population
synthesis models to their spectro-photometric SEDs we determine their ages to
be $2.4^{+0.8}_{-0.6}$~Gyr and $1.7\pm0.3$~Gyr, respectively, which implies that the two galaxies have higher mass-to-light ratios than most quiescent $z\sim2$ galaxies in other studies.
We find no direct evidence for active star-formation or AGN activity in either of the two
galaxies, based on the non-detection of emission lines. Based on the derived
redshifts and stellar ages we estimate the formation redshifts to be
$z=4.3^{+3.4}_{-1.2}$ and  $z=4.3^{+1.0}_{-0.6}$, respectively. We use the
increased spatial resolution due to the gravitational lensing to derive constraints on the
morphology. Fitting Sersic profiles to the de-lensed images of the two galaxies confirms their compactness, with one
of them being spheroid-like, and the other providing the first confirmation of a passive lenticular galaxy at a spectroscopically derived redshift $z\sim2$.
\end{abstract}


\keywords{galaxies: fundamental parameters, galaxies: high-redshift, galaxies: stellar content, galaxies: formation}



\section{Introduction}
\label{Introduction_Section}

The study of galaxy formation and evolution has taken giant leaps forward in the last 15 years. In particular the use of the Lyman-break (or drop-out) selection technique has revealed large numbers of star-forming galaxies over a wide range of redshifts all the way back to $z\approx10$ \citep[e.g.,][]{2003ApJ...592..728S,2011Natur.469..504B}.  
The current overall stellar mass density is about 5$\times$10$^8$ M$_{\sun}$ Mpc$^{-3}$ \citep{1998ApJ...503..518F,2003ApJ...587...25D, 2003ApJS..149..289B, 2012MNRAS.421..621B} and the early build-up of this mass is now becoming observationally accessible (only accurate to a factor of  a few at the highest redshifts) from roughly $z=9$ (0.1\%  of the present value) to $z=2.5$ (10--20\% of the present value) \citep{2010ApJ...716L.103L}. Roughly 80--90\% of present day stars must have formed at $z<2.5$ \citep{2009ApJ...701.1765M}. The cosmic star-formation history at $z<2.5$ seems to proceed in a manner that has been referred to as "Downsizing", i.e.  proceeding from high- to low-mass systems as a function of cosmic time \citep[e.g.][]{2005ApJ...619L.135J}. A similar picture emerges from the morphological evolution of galaxies at different stellar masses \citep{2011ApJ...743..146C}.
\\It has become evident that the redshift interval $1.5<z<4$ constitutes the most important phase in the formation and evolution of massive ($\gtrsim 10^{11}$ M$_{\sun}$) galaxies. This is the cosmic era when massive galaxies had the peak of their star-formation and AGN activity and when their morphologies began to be transformed from being disk-like into being dominated by spheroids, and even having their star formation quenched \citep{1996ApJ...460L...1L, 1996MNRAS.283.1388M, 1998ApJ...498..106M, 2004ApJ...615..209H,2005AAS...207.7804S, 2006ApJ...651..142H,2007MNRAS.377.1717K, 2008ApJ...680..224Z,2012A&A...537L...8C}. These processes are probably connected, with massive star-forming galaxies building up their stellar mass, subsequent mergers igniting nuclear starbursts, transforming the morphology from disks to spheroids, and finally ignition of the AGN shutting down star-formation.
Furthermore, it appears that the mass-metallicity and color-magnitude relations we observe at lower redshifts were most likely established in that era \citep{2010MNRAS.408.2115M}.
\\The advent of deep near-infrared (NIR) surveys revealed a population of massive $z\sim2$ galaxies that seem to have had their star-formation quenched already at $z\gtrsim3$ \citep{2003ApJ...587L..79F,2003ApJ...587L..83V}. These must have been the first massive galaxies that stopped forming stars.
Many of these galaxies have been shown to be extremely compact (effective radii $r_e\sim1$~kpc), with mean sizes of about one sixth to one third 
of that of local elliptical galaxies of the same stellar mass \citep{2006ApJ...650...18T,2006MNRAS.373L..36T,2007ApJ...671..285T,2007ApJ...656...66Z,2008ApJ...687L..61B,2009ApJ...700..221K,2009ApJ...705..255T,2012ApJ...746..162N}.
This probably reflects their very early formation epochs, when the Universe was much denser than today.
\cite{2009ApJ...692L.118T} and \cite{2010ApJ...720..723T} find that in the local Universe this type of galaxies is extremely rare, which means that the massive compact galaxies we see at $z\sim2$ must have experienced significant structural evolution since then. The picture that has been put together in the last few years is that of inside-out growth due to mostly minor merging \citep{2009ApJ...697.1290B,2009ApJ...699L.178N,2010ApJ...709.1018V,2012ApJ...746..162N,2013MNRAS.428..641O}.
Quiescent galaxies contain a large fraction of the total stellar mass at $z\sim2$, and need to be studied in greater detail in order to obtain a complete picture of all types of galaxy populations at $z\sim2$.
\\Our understanding of these distant massive quiescent galaxies is still limited, as redshifts
and stellar population studies have been mainly based on broad-band
photometry. Spectroscopic confirmation of their redshifts and stellar
properties is very important but also very challenging, as these objects are extremely
faint in the observed optical wavelength range ($I>25.5$mag) and in the absence of star formation do not exhibit emission lines.
Up to now spectroscopic investigations have only been possible for the brightest galaxies of this class which may not be representative for the overall population.
Previous spectroscopic studies, such as GMASS
\citep{2008A&A...482...21C,2008ASPC..381..303K}, have used $\sim32$ hours of 8-m
telescope time (per MOS mask) to obtain redshifts. To push from mere redshift determinations to stellar population studies and stellar velocity dispersions, higher signal-to-noise ratios are required, which reduces the target sample size even further and renders statistical studies challenging with currently available instrumentation.
\\As a way to circumvent this problem we select massive quiescent galaxies that have been strongly lensed by intermediate redshift $(0.1<z<0.8)$ galaxy clusters.
Gravitational lensing can boost the observed magnitudes to a level at which spectroscopy of continuum and absorption lines becomes feasible within a reasonable amount of observing time, even for galaxies with luminosities around $L\star$.
The approach to make use of the strong gravitational lensing effect of galaxy clusters to identify galaxies at high redshifts \citep{1998Ap&SS.263...55P,2004ASSL..301...27R} for easier photometric \citep{1999ASPC..191..241P, 2007HiA....14..250P} and spectroscopic follow-up studies \citep{2001A&A...378..394C, 2003A&A...397..839L} has been in use for more than a decade and has been proven a very powerful method for many types of high-redshift objects \citep{2006A&A...456..861R,2007A&A...469...47S,2008A&A...477...55H,2010A&A...509A.105M,2011MNRAS.413..643R,2012MNRAS.427.1973C,2012MNRAS.427.1953C}.
We now extend this method to the above-described population of distant massive quiescent galaxies.
With the help of already available mass-models \citep{2007ApJ...661L..33E,2010MNRAS.404..325R} for clusters in which strong lensing is observed, we can determine the intrinsic
luminosities of distant massive quiescent galaxies identified in the high-magnification regions of the cluster images. \\In this way we have identified significantly magnified distant massive quiescent galaxies with intrinsic luminosities (assuming $z\approx2$) around $L\star$ with observed $Ks$-band (Vega) magnitudes $\lesssim19$mag. 
These galaxies are within reach of spectrographs like X-Shooter on the VLT with total exposure times of 5--10 h, as has been shown in recent studies \citep{2011ApJ...736L...9V,2012arXiv1211.3424V,2012ApJ...754....3T}.
\\In this paper we present NIR spectra of two gravitationally-lensed distant massive quiescent galaxies obtained with VLT/X-Shooter, combined with broad-band photometry from the {\it Hubble Space Telescope (HST)} and ground-based facilities. In the next section we describe how the targets were selected from imaging data, and how the spectroscopic data were reduced. Sect.~\ref{Analysis_Section} contains the analysis of the spectra, with the obtained redshifts, stellar population synthesis models, constraints on potentially on-going star formation, and a structural analysis for the galaxies with resolved imaging from HST. In Sect.~\ref{Discussion_Section} we discuss the interpretation of these results with respect to currently discussed paradigms of galaxy formation and evolution.
We adopt a flat $\Lambda CDM$ cosmology with $H_{0}=70.4~km~s^{-1}~Mpc^{-1}$, $\Omega_{M}=0.272$ and
$\Omega_{\Lambda}=0.728$, according to \citet{2011ApJS..192...18K}. Magnitudes and colors in this paper are given in the Vega magnitude system, except where they are explicitly marked to be in the AB system.

\section{Object selection, data reduction and SEDs}
\label{Data_Section}

\subsection{Target selection}
\label{Target selection_Section}

The two candidate $z\sim2$ massive quiescent galaxies which we present in this article were identified behind the strong lensing clusters Abell~1413 \citep[$z=0.14$,][]{2000ApJS..129..435B} and
MACS2129-0741 \citep[$z=0.59$,][]{2007ApJ...661L..33E}. For A1413 $J$-
and $Ks$-band images were available from the WIRC instrument \citep{2003SPIE.4841..451W} on the Palomar 200-inch telescope, with $5\sigma$ depths of 22.1mag in $J$ and
20.5mag in $Ks$, respectively \citep[see also][for details on the data reduction]{2010MNRAS.404..325R}. Near-Infrared imaging data of MACS2129-0741 was available from VLT/ISAAC
(with $5\sigma$ depths of 22.9mag and 20.8mag in $J$ and $Ks$, respectively). We adopt $MAG AUTO$ in SExtractor \citep{1996A&AS..117..393B} for total magnitudes and determine  the photometric zeropoints using bright ($<15$mag in each band) stars from the 2MASS catalog \citep{2006AJ....131.1163S} within the same field.
For both clusters, $J-K_{s}$ colors were obtained by using the dual mode of SExtractor with the $Ks$-band image as detection image and measuring fluxes in small circular apertures of about the size of the seeing disc.
To account for seeing differences, the WIRC $J$-band image of A1413 was Gaussian-convolved to the $K_{s}$-band seeing.
In both the ISAAC images of MACS2129-0741, we measure seeing FWHMs of $0\farcs5$, thus no convolution was necessary.
Galactic extinction corrections were applied according to the \citet{1998ApJ...500..525S} extinction maps.

One of the red galaxies, which we identified in the A1413 field, and which we
refer to as A1413-1, was selected as one of the spectroscopic targets. Out of
the lensed sources identified in the field of MACS2129-0741, we present spectroscopy for
the brightest one which we refer to as MACS2129-1. Both were selected mainly for their relatively bright apparent magnitudes and the lensing magnification.
The color-criterion of $J_{s}-K_{s}>2.3$, originally proposed by \citet{2003ApJ...587L..79F} was used as a guideline, but given that the brightest lensed red galaxy in the field of A1413 exhibits a color of $J-K_{s}\sim2.15$, this was not followed strictly.
A short summary of the two objects is given in Table \ref{table1}.

At the time of target selection we assumed a redshift of the sources of
$z\sim2$ and using the mass models from \citet{2010MNRAS.404..325R}, we
estimated the lensing magnifications to be 0.8mag for A1413-1 and 1.87mag for
MACS2129-1, respectively. Later we will refine the de-lensing, based on more
accurate redshifts, to be described in the following chapters.

\begin{deluxetable}{@{}ccccccc@{}}
\tablecaption{Coordinates, magnitudes and colors of the two target galaxies. $^{(a)}$ Observed magnitude. $^{(b)}$ Estimated intrinsic magnitude inferred from the lensing
magnification at $z\approx2$.\label{table1}}
\tablewidth{0pt}
\tablehead{Name & Facility & $Ks_{obs}^{(a)}$ & $J-Ks$ & $Ks_{int}^{(b)}$ & RA & DEC}
\startdata
\hline
A1413-1 & WIRC & $19.01\pm0.06$ & 2.15 & 19.81 & $11^{h}55^{m}15.4^{s}$ & $+23^{\circ}24'52\farcs7'$ \\
\hline
MACS2129-1 & ISAAC & $17.81\pm0.01$ & 2.42 & 19.68 & $21^{h}29^{m}22.2^{s}$ & $-07^{\circ}41'31\farcs2'$ \\  
\hline
\enddata 
\end{deluxetable}

\begin{figure}[h!]
\includegraphics[angle=0,scale=0.688,width=1.0\linewidth]{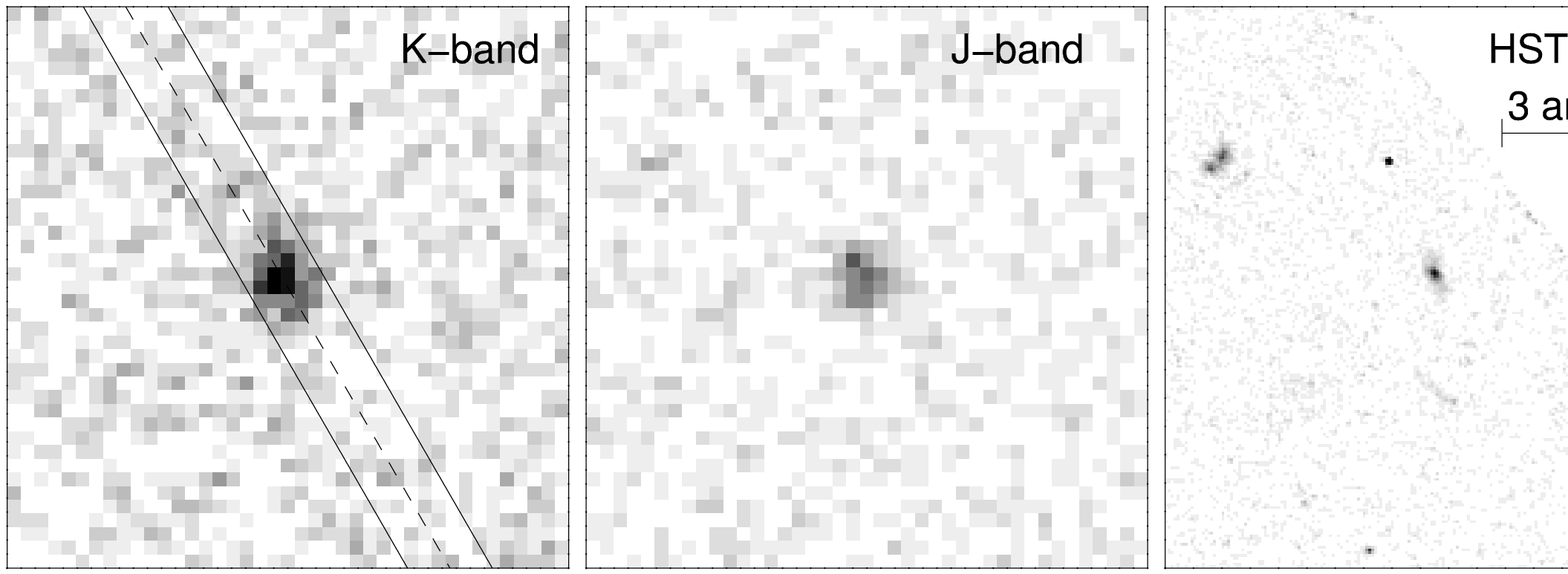}
\includegraphics[angle=0,scale=0.688,width=1.0\linewidth]{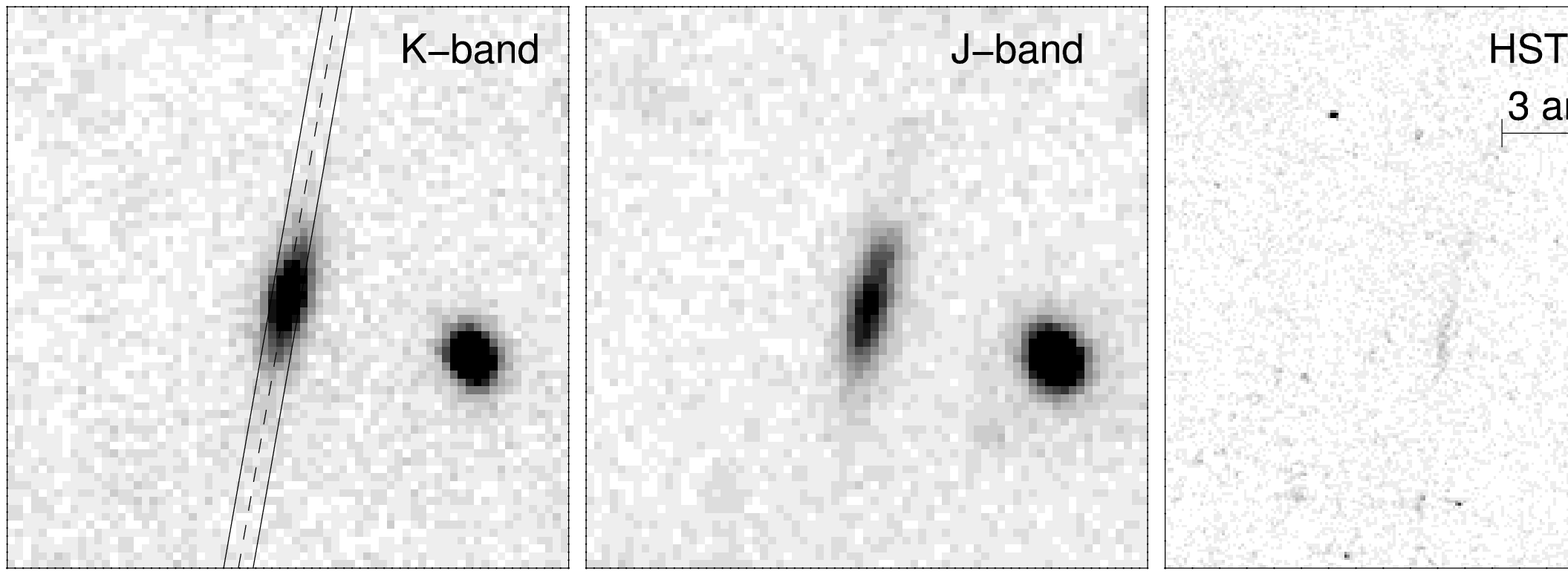}
\caption{Upper panel: Cutout images of the gravitationally-lensed galaxy A1413-1 in the (WIRC) $Ks$- and $J$-bands, and in the HST/ACS F775W band. Lower panel: Cutout images of the gravitationally-lensed galaxy MACS2129-1 in the (ISAAC) $Ks$- and $J$-bands, and in the HST/ACS F606W band. \label{pic_macs2129a1413}}
\end{figure}

In Fig. \ref{pic_macs2129a1413} we show cutout images of the two galaxies.
It is clearly visible that the image of MACS2129-1 is stretched to an arc with an extension of about $1\farcs6$ (FWHM) by the strong lensing effect of the foreground cluster.
We will discuss its morphology later in more detail, where MACS2129-1 turns out to be a disk-like galaxy in the source-plane at $z\sim2$.

\subsection{Spectroscopic data}
\label{Spectro_Section}

Both targets (A1413-1 and MACS2129-1) were observed as part of the ESO program
087.B-0812 (PI: Toft) with the X-shooter spectrograph \citep{2006SPIE.6269E..98D} on
VLT/UT2.  This instrument is a medium-resolution Echelle spectrograph capable of obtaining spectra from the UV ($\sim300nm$) to the NIR ($\sim2500nm$) simultaneously. The data were collected with a generic offset template that obtained six 480s long exposures per observation block in the NIR arm, using the $0\farcs9$ slit. The total exposure times in the NIR arm were 3.2h for both A1413-1 and MACS2129-1, in seeing conditions which varied between $0\farcs6$ and $1\farcs5$.

\subsection{Spectroscopic Data Reduction}
\label{Spectro_Reduction_Section}

For the reduction of the spectroscopic data the ESO X-shooter pipeline version
1.3.7 \citep{2010SPIE.7737E..56M} was used. The calibration steps (master
darks, order prediction, flat fields, and the 2d maps for later rectification
of the spectra) were run for each night separately, with the default parameters
in the pipeline \citep{2011AN....332..227G}.

With the output from these five calibration steps the scientific raw frames are reduced: The Echelle spectra get dark-subtracted, flatfielded, and rectified. We experiment with different pipeline recipes and parameter
settings for the reduction of the science frames. It turned out that reducing
them in "stare-mode", where the sky-background for a given object frame is
estimated from that same object frame, results in poor sky-subtraction and large skyline-residuals,
especially in the wavelength range of the $K$-band. Instead, using the nodding
recipe, where dithered frames are used pairwise for subtraction of the
sky-background, provides the best sky-subtraction and signal-to-noise ratio (SNR) for
data from the NIR arm. Thus we decided to use this method for the
reduction of the object frames, which results in one reduced 2d image for each
adjacent pair of raw object frames, which comprises 16min of exposure time. For consistency, the data from the UV and optical arms were also reduced with the nodding recipe and sampled onto the same grid.

\subsection{Flux calibration, combination and 1d extraction}
\label{Spectro_Calibration_Section}

For flux calibration, standard stars were observed in each night in which the
data were obtained. They are taken from a list of eight white dwarfs, the
intrinsic spectra of which have been modeled very accurately. Each standard star
observation is reduced with the same calibration data as the science frames
from the same night and the same pipeline parameters are employed for the
reduction of the raw standard frames. From the observed shape of the flux
standard spectrum we derive the spectral response curve: the 1d extracted
spectrum of the standard star is divided by the tabulated known intrinsic
spectrum (which is first interpolated and re-gridded to the same pixel binning
on the wavelength axis) and the resulting curve is smoothed with a kernel of 15 pixels to obtain a smooth response curve and to minimize
artificial effects introduced by pixels with abnormal values. We apply the flux
calibration by first dividing the 2d spectra of our science targets as well as
the 1d response curve by the respective exposure times (thus normalizing them
to 1s exposure time) and then dividing the normalized 2d spectra along the wavelength axis by
the normalized response curve for the respective night.  In the NIR, atmospheric extinction is negligible, and no atmospheric extinction table is
available for that wavelength range in the pipeline release.
Telluric absorption is accounted for by the flux standard, as we apply only little smoothing to the overall response curve.

By collapsing and weighting the spectrum along the wavelength axis, we
determine the shapes of the spectral point spread function (SPSF) on the
individual 2d spectra which we get as output from the pipeline, and thus
determine the shifts to register and co-add the 2d spectra, which is done as a mean combination with rejection of outliers, which we define as pixel values which deviate from the mean by more than $10\sigma$. We apply the same calibrations to the error maps as well, and combine them according to
\begin{equation}
Err_{combined}=\frac{\sqrt{\sum_{i=1}^n Err_{i}^{2}}}{n}
\end{equation}
The final SPSF is determined by collapsing the weighted 2d stack along the wavelength axis in the $H$-band wavelength range (where the SNR per pixel is highest) and the 1d spectrum is extracted by
applying the corresponding normalized weights, a procedure which is similar to the optimal extraction procedure described in \cite{1986PASP...98..609H}. The extraction window spans a spatial extension of $1\farcs8$ around the  centre of the trace for the point-source-like A1413-1, and for MACS2129-1, which is extended to an arc with a FWHM of $1\farcs6$, we extract the 1d spectrum in a $2\farcs6$ window. A 1d error spectrum is extracted the following way: with NSP being the normalized SPSF along the extraction window, the 1d variance is calculated as
\begin{equation}
Var(\lambda)=\frac{\sum_{y=y1}^{y2}Err(\lambda,y)^{2}*NSP(y)^{2}}{(\sum_{y=y1}^{y2} NSP(y)^2)^2}
\end{equation}
and the 1d errors are then the square root of that variance.

Galactic extinction corrections are taken from \citet{1998ApJ...500..525S} and implemented with the $fm\_unred$ code in IDL.

To check the flux level we integrate the extracted 1d spectra over the transmission curves of the $Ks$-band filters of the instrument from which the imaging data originate and compare with the photometric measurement. We subsequently scale both 1d spectra to match the $Ks$-band photometry, thus correcting for slit losses.

\subsection{HST imaging data}
\label{HST_data_Section}

To add further information to the SEDs of the two galaxies we include photometry from publicly available {\it HST} images of the two galaxies.
For A1413-1 we use images obtained with the Advanced Camera for Surveys (ACS) in the $F775W$ and $F850LP$ bands as part of program 9292 (PI: Holland Ford).
For MACS2129-1 we use images obtained with ACS in the $F606W$ band and with Wide Field Camera 3 (WFC3) in the $F105W$, $F110W$, $F125W$, $F140W$ and $F160W$ bands. These images were obtained as part of the CLASH Survey (program 12100, PI: Marc Postman).
Photometry for A1413-1 is done with circular apertures (diameter $2\arcsec$).
For MACS2129-1, which is extended to an arc, we use SExtractor \citep{1996A&AS..117..393B} to define elliptical apertures and measure the total counts using $MAG AUTO$. For the $F606W$ image of MAC2129-1 we used the $F110W$ image (re-binned and rotated to the $F606W$ pixel size and orientation) to define the aperture as the source is very faint in this band. We adopt zeropoints and aperture corrections from the ACS and WFC3 instrument handbooks. The \citet{1998ApJ...500..525S} maps again provide the galactic extinction corrections.
The error bars on the photometry are conservative estimations, in order to account for cross-calibration issues when used together with the X-Shooter spectra to construct the SEDs of the two galaxies.
We check the accuracy of the relative flux-calibration in both spectra by over-plotting the HST photometry and conclude from the good agreement that no further flux correction of the spectra is necessary.

\subsection{Spitzer imaging data}
\label{MIPS_data_Section}

Both MACS2129 and A1413 had also been observed with the Multiband Imaging Photometer (MIPS) on the Spitzer Space Telescope, in the $24\mu$-band. 1320s of integration time were obtained on MACS2129 as part of program 50610 (PI: Yun) and 480s on A1413 as part of program 41011 (PI: Egami). Both galaxies in our study are undetected on these images. Because of noise structures in the images it was also not possible to derive upper limits on their fluxes.

\section{Data analysis}
\label{Analysis_Section}

The signal-to-noise ratio in the spectra of both targeted galaxies is relatively low. We estimated the S/N in bins of several nm along the wavelength axis, where we sum up the flux in the object trace in a spatial window of about one seeing FWHM, and estimated the noise from the regions of the same size without object flux below and above the trace. For MACS2129-1 we thus estimate a mean $S/N/\mathrm{\AA}$ of $\sim1.3$, $\sim2.5$, and $\sim2.1$ in the $J$-, $H$-, and $Ks$-band wavelength regions, respectively. For the spectrum of the significantly fainter A1413-1 the values are $\sim0.9$, $\sim1.2$, and $\sim1.0$ for $J$, $H$, and $Ks$, respectively.

\subsection{Redshift determination and spectral energy distribution analysis}
\subsubsection{Emission and absorption lines}
\label{Lines_Section}

The NIR 2d spectra of both galaxies were examined visually to search for
emission lines and absorption features. Visualizations of both the NIR 2d and 1d spectra are shown in
Figures \ref{spec_a1413}, \ref{spec_macs2129}, and \ref{specs_halpha}. For better visibility, we also smooth
them along the wavelength axis. We do not detect emission lines in the
spectra of any of the two galaxies. At several positions we see 
hints of absorption lines, but we did not consider these significant and reliable 
enough for an independent robust redshift measurement. Instead we proceed with an analysis of the spectral energy distributions and photometric redshift measurements in Sect.~\ref{SED_Section}
and Sect.~\ref{LePhare_Section}. Guided by those we then in Sect.~\ref{tentative} return to the issue of the nature of the 
tentative spectral lines.

We do not find any trace in the UV arm data for both objects. Over the whole
optical wavelength range, a trace is detected from MACS2129-1, and partly from
A1413-1.  Even in big bins, however, we only find very low S/N. As a
consequence we will later rely on the HST broad-band magnitudes for this
wavelength range.

\subsubsection{Construction of spectro-photometric Spectral Energy Distributions}
\label{SED_Section}

Although the potential absorption lines in the NIR spectra of both galaxies are not significant enough to reliably determine independent spectroscopic redshifts, 
there is still the possibility to determine their redshifts
(although with higher uncertainties than for line measurements) from the shape
of the NIR spectrum, i.e. mainly the position of the Balmer/4000\AA-break
which both are distinctive (but partly overlapping) features in stellar populations with ages of several
hundred Myrs and above.

In order to have a well-sampled Spectral Energy Distribution (SED) with
sufficient S/N in each data point, we bin the NIR spectrum, with bin sizes
ranging between 15nm and 70nm.

For each bin we regard the wavelength range as the transmission curve $T_{synth}(\lambda)$ of a
``synthetic filter'' with cut-on and cut-off wavelengths $\lambda_{1}$ and
$\lambda_{2}$ and mean wavelength $\lambda_{mean}$ , which we all define below.
Furthermore, the combined fluxes and errors in each bin are calculated from the
original fluxes and errors as follows:  \\
\begin{equation}
weight(\lambda)=\frac{1}{err(\lambda)^2}
\end{equation}
\begin{equation}
T_{synth}(\lambda)=\frac{weight(\lambda)}{max(weight(\lambda))}
\end{equation}
\begin{equation}
\lambda_{mean}=\frac{\int_{\lambda1}^{\lambda2}\lambda\cdot T_{synth}(\lambda)\cdot d\lambda}{\int_{\lambda1}^{\lambda2}T_{synth}(\lambda)\cdot d\lambda}
\end{equation}
\begin{equation}
F_{\lambda,bin}=\frac{\sum_{\lambda1}^{\lambda2}{F_{\lambda}(\lambda)\cdot T_{synth}(\lambda)}}{\sum_{\lambda1}^{\lambda2}T_{synth}(\lambda)}
\end{equation}
\begin{equation}
Err_{bin}=\frac{\sqrt{\sum_{\lambda1}^{\lambda2}{Err(\lambda)^2\cdot T_{synth}(\lambda)^2}}}{\sum_{\lambda1}^{\lambda2}T_{synth}(\lambda)}
\end{equation}

One can see from the above formulae that the relative weights of each
contributing wavelength pixel are used as ``transmission'' of our individually
defined ``synthetic filters''.

To obtain reliable constraints on the redshifts via the position of the
Balmer/4000\AA-break we want to sample this region with several data points on
each side of the feature. From visual inspection of the shape of both spectra
we conclude that the Balmer/4000\AA-break is located in the range of the $Y$- and
$J$-bands. We thus manually define 10 bins in this wavelength region, which
turn out to exhibit S/N ratios between 5.5 and 16 for the A1413-1 spectrum and
between 13 and 32 for the MACS2129-1 spectrum. We then divide the $H$-band
wavelength region into two parts, one ranging from 1440nm to 1521nm with lower
S/N, and one from 1521nm to 1798nm, where the S/N ratio reaches its highest
values across the whole NIR wavelength range. Both of them are binned by an
algorithm which sets the bin sizes such that they result in a certain S/N
ratio. We thus bin the short wavelength region of the H-band with S/N per bin
of 20 for the A1413-1 spectrum and 50 for MACS2129-1, and for the longer
wavelength range of the $H$-band we bin up to a S/N per bin of 35 for A1413-1 and
75 for MACS2129-1. For the $Ks$-band wavelength range from 1970nm to 2300nm we
again define bins manually: we exclude noisy regions (where absorption is high)
and split the rest into 7 bins, with S/N ratios between 6 and 24 for A1413-1
and 22 and 49 for MACS2129-1. Wavelengths beyond the end of the $Ks$-band
at $2.3\mu m$ were excluded as the sky-subtraction here was not 
sufficiently good to allow robust measurements.

In addition to the SEDs constructed out of the NIR spectra, we add
the above-mentioned HST photometry in the $F606W$ band for MACS2129-1 and in 
the $F775W$ and $F850LP$ bands for A1413-1, in order to extend the SEDs of 
the two galaxies also to optical wavelengths.

A complete overview of the SEDs is given in Tables \ref{tab_a1413-1} and \ref{tab_macs2129-1}, where we include the quasi-photometric magnitudes of the binned NIR spectra, and the broad-band magnitudes from the HST images described in section~\ref{HST_data_Section}.

\subsubsection{Stellar population synthesis fits to the SEDs}
\label{LePhare_Section}

\begin{figure}[h!]
\vspace{0.5cm}
\includegraphics[angle=90,scale=1.0,width=1.0\linewidth]{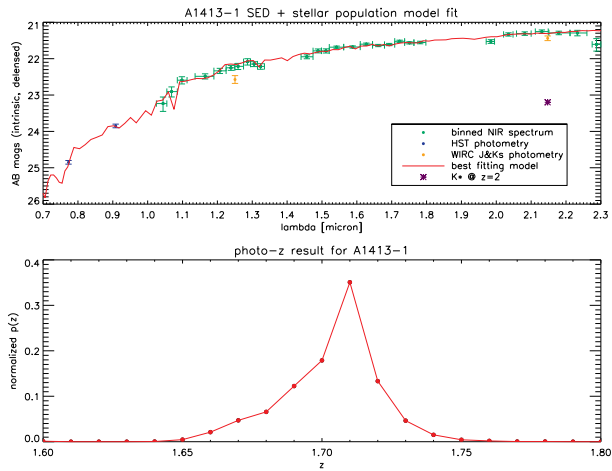}
\caption{Upper panel: Spectro-photometric SED of the gravitationally-lensed galaxy A1413-1, together with HST/ACS and Palomar/WIRC photometry. Overplotted is the best-fit BC03 model, a 2.4~Gyr passively evolving population with little dust and a stellar mass of $M_{\star}=7.6\cdot10^{10}M_{\odot}$. Lower Panel: Probability distribution of the photometric redshift for A1413-1. \label{afig}}
\end{figure}

\begin{figure}[h!]
\vspace{0.5cm}
\includegraphics[angle=90,scale=1.0,width=1.0\linewidth]{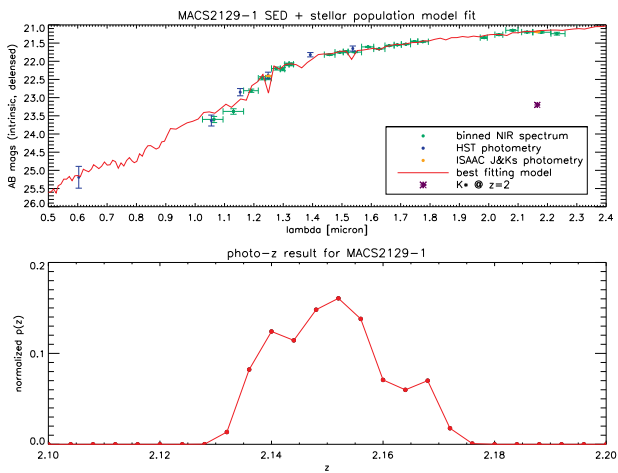}
\caption{Upper panel: Spectro-photometric SED of the gravitationally-lensed galaxy MACS2129-1. Overplotted in red is the best-fit BC03 model with a Chabrier IMF, a 1.7~Gyr population with a stellar mass of $M_{\star}=1.1\cdot10^{11}M_{\odot}$. Lower panel: Probability distribution of the photometric redshift for MACS2129-1. \label{macsfig}}
\end{figure}

We use the multi-wavelength SEDs, described in section~\ref{SED_Section}, 
to fit stellar population synthesis models, in order to determine photometric
redshifts as well as stellar ages, masses, and dust extinctions from them.
Given the limited SNR of the available spectra and the low significance of absorption lines, it is not possible to place robust constraints on the 
metallicities.  We fit the data using the LePhare code
\citep{1999MNRAS.310..540A,2006A&A...457..841I} and galaxy models from
\citet[][BC03 hereafter]{2003MNRAS.344.1000B}, based on the Chabrier initial mass function (IMF)
\citep{2003PASP..115..763C}, and the Calzetti extinction law \citep{1994ApJ...429..582C,2000ApJ...533..682C}. The assumed star formation histories in the BC03 models follow an exponential declining rate, $SFR\propto
e^{-\frac{t}{\tau}}$ , with 9 different e-folding timescales $\tau$, ranging
from 0.1~Gyr to 30~Gyr. Furthermore, they come with three different
metallicities, $Z=0.02$ (solar value), $Z=0.008$ and $Z=0.004$. LePhare is
based on a $\chi^{2}$ template-fitting procedure
\citep{1999MNRAS.310..540A,2002MNRAS.329..355A}, with an input grid comprising
of the above-mentioned list of 27 BC03 models, a range of redshifts $z$ in steps
$\Delta z$, extinction coefficients $E_{B-V}$ , and a list of ages for the
models. A library of theoretical magnitudes is built by redshifting each SED in
steps of $\Delta z$ and convolving them with the transmission curves of the
filters. In our case, these filters are those defined synthetically by the
binning procedure described in Sec.~\ref{SED_Section}. LePhare also takes into account the
opacity of the inter-galactic medium as described in \citet[][see also \citealt{1990A&A...228..299M}]{1995ApJ...441...18M}.
After the determination of the $z$ value taken from the input grid which minimizes the merit function $\chi^{2}$, the best fitting
redshift is derived with a parabolic interpolation of the redshift probability
distribution. Following \citet{2006A&A...457..841I,2009ApJ...690.1236I} the
$1\sigma$ level errors obtained from the probability distribution
function are a reliable estimate of the actual photo-z accuracy.  In the
computation of the library of theoretical galaxy magnitudes, we follow an
iterative approach: we first adopt a wide-spaced grid covering a large range of
input values to obtain first estimates on the (photometric) redshift, the range
of possible stellar ages, and the extinction coefficient $E_{B-V}$. With the
help of those, we narrow down the range of redshifts, ages and $E_{B-V}$ values
used in a second iteration of the LePhare fitting procedure, enabling us to use
smaller steps and thus a finer input grid for $z$ and $E_{B-V}$. The main
motivation for this approach is the technical limitation of the libraries to a
maximum of $9\cdot10^{5}$ entries, i.e. the product of number of input BC03
models, extinction laws, and the numbers of entries on the grids of ages,
$\Delta z$ steps and $E_{B-V}$ values. The finer sampling on the redshift grid
allows us to reduce the systematics and obtain more detailed insights in the
shape of the redshift probability distribution, especially about potential
secondary or double peaks. The lower uncertainties in the redshift due to the
finer redshift grid and the smaller steps in the used $E_{B-V}$ values allow
us to obtain a more accurate estimation of the latter and thus result in more
reliable constraints on the stellar ages. The best fit models for both
galaxies are overplotted in Fig. \ref{afig} and Fig. \ref{macsfig}, and the
results for the derived physical parameters are summarized in Table
\ref{tab_fit}. It has to be emphasized here that the stellar masses are
subject to systematic uncertainties (which can easily exceed a factor of 2) which are not included in the quoted error
budget but originate in the uncertainty on the initial mass function (IMF)
which is not well constrained at high redshifts.
It turns out that the input SEDs are best fit by the BC03 models with an e-folding timescale for star formation of 0.1~Gyr (A1413-1) and 0.3~Gyr (MACS2129-1), respectively.
The model fit for A1413-1 indicates no on-going star formation (with an upper limit of $\sim0.2~M_{\odot}yr^{-1}$), and for MACS2129-1 the fit allows for a star formation rate of $2\pm2~M_{\odot}yr^{-1}$ which according to the resulting specific star formation rate of $(1.8\pm1.8)\cdot10^{-11}yr^{-1}$ means that MACS2129-1 can be regarded as a passively evolving galaxy. We emphasize that this value for star formation is solely based on the best-fit model with the assumption of an exponentially declining star formation rate. This approximation does most likely not represent reality \citep{2012arXiv1212.4153A} and as a consequence the computed star formation rates should not be taken too literally. Given the non-detection of emission lines we conclude that the MACS2129-1 spectrum is still consistent with no ongoing star formation.

\begin{deluxetable}{@{}ccccccc@{}}
\tablecaption{LePhare fitting results for the lensed galaxies A1413-1 and MACS2129-1 \label{tab_fit}}
\tablewidth{0pt}
\tablehead{
Parameter & A1413-1& MACS2129-1 
}

\startdata

$z_{phot}$ & $1.71\pm0.02$ & $2.15\pm0.01$ \\ 
Age~[Gyr] & $2.4^{+0.8}_{-0.6}$ &  $1.7\pm0.3$  \\ 
$E_{B-V}$~[mag] & $0.02\pm0.01$ & $0.1\pm0.02$ \\ 
$A_{V}$~[mag] & $0.08\pm0.08$ & $0.41\pm0.28$ \\
$M_{\star}$~[$10^{10}~M_{\odot}$]  & $7.6\pm1.1$ & $11.0\pm2.8$ \\ 
SFR~[$M_{\odot}yr^{-1}$] & $0$ & $2\pm2$ \\
sSFR~ [$yr^{-1}$]  & $0$ & $(1.8\pm1.8)\cdot10^{-11}$ \\
$z_{form}$ & $4.3^{+3.4}_{-1.2}$ & $4.3^{+1.0}_{-0.6}$ \\
\hline
\enddata
\end{deluxetable}

\subsubsection{Nature of tentatively detected spectral lines}
\label{tentative}

Guided by the photometric redshifts found above we now return to the issue of tentative 
absorption lines in the spectrum. As can be seen in Fig.~\ref{spec_a1413} and Fig.~\ref{spec_macs2129} 
there are indications of very broad H$\beta$ absorption at the expected positions dictated by 
the photometric redshifts for both MACS2129-1 ($S/N\sim3.5$) and A1413. For MACS2129-1 there are also hints
of absorption at the expected position of CaII + H$\delta$. Based on the position of the 
tentative H$\beta$ lines we infer spectroscopic redshifts of $z=2.1477\pm0.0007$ and 
$z=1.707\pm0.002$, which are fully consistent with the photometric redshifts. We defer further analysis of these tentative features (e.g. velocity dispersions) to a future
study.
To double-check the SED fitting performed in Sect. \ref{LePhare_Section} we re-ran the same procedure, but this time with the redshifts fixed to the spectroscopic redshifts inferred from the H$\beta$ lines. The results from the new LePhare fits show no significant changes with respect to the results reported in Table \ref{tab_fit}.

\subsection{Profile fits to the de-lensed images}
\label{GALFIT_Section}

\begin{figure}[t!]
\includegraphics[angle=0,scale=0.75,width=1.0\linewidth]{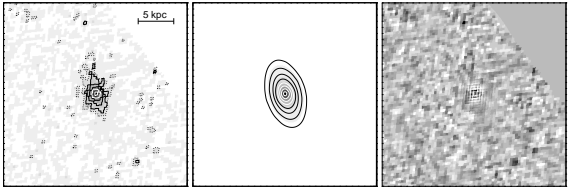}
\caption{Left panel: The de-lensed image of the gravitationally-lensed distant massive quiescent galaxy A1413-1, reconstructed from the HST F850LP image, based on the derived redshift of $z=1.71$. Middle panel: A Sersic profile fit to the de-lensed image, obtained with GALFIT. For further information about the fit results, see Table \ref{Allisons}. Right panel: Residuals from the Sersic fit. 
\label{JR_fig_A}}
\end{figure}

\begin{figure}[t!]
\includegraphics[angle=0,scale=0.75,width=1.0\linewidth]{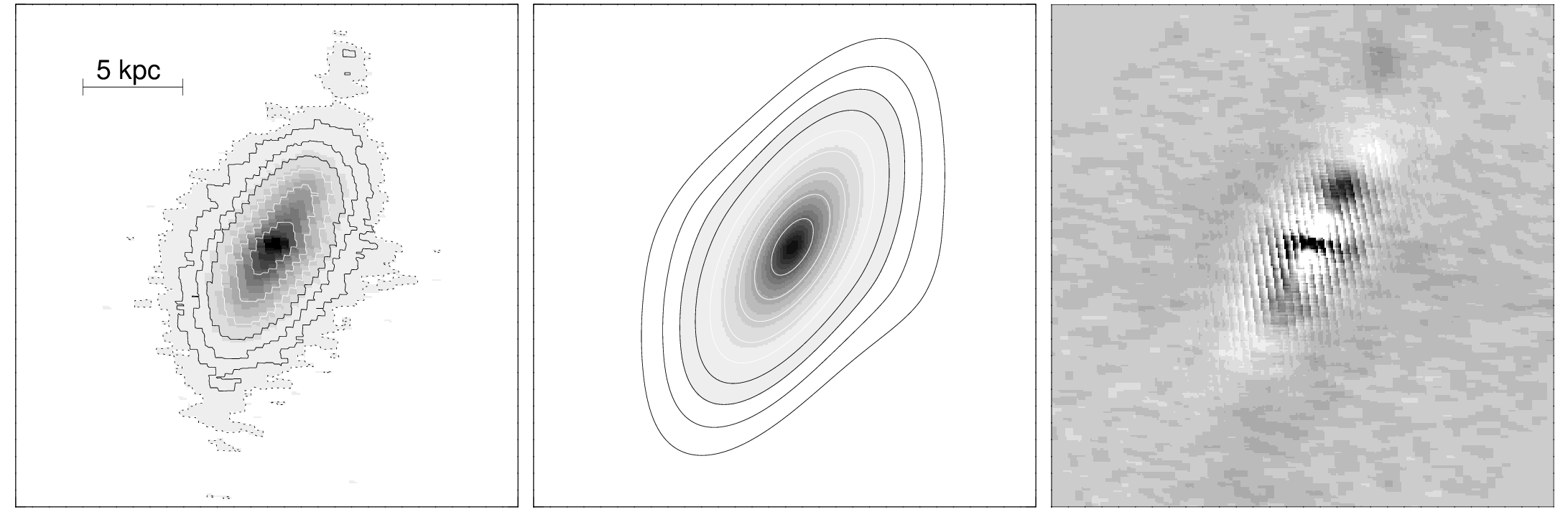}
\includegraphics[angle=0,scale=0.75,width=1.0\linewidth]{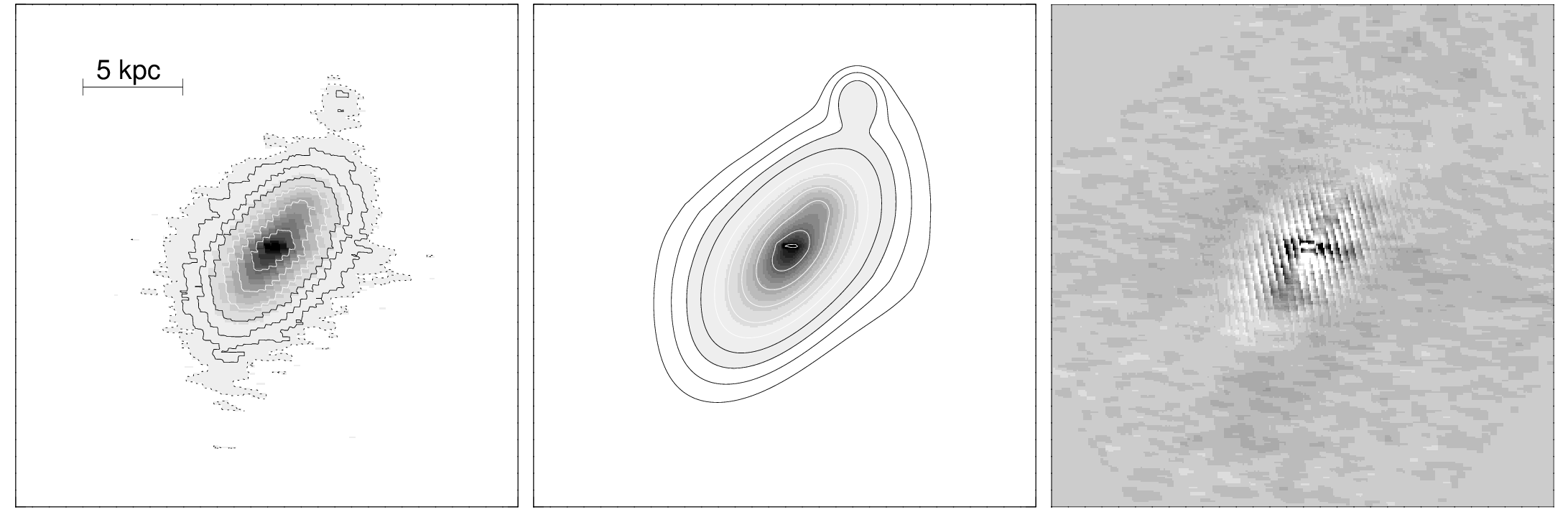}
\caption{Left panel: The de-lensed image of the gravitationally-lensed distant massive quiescent galaxy MACS2129-1, reconstructed from the HST F160W image, based on the derived redshift of $z=2.15$. Middle panel: A Sersic profile fit to the de-lensed image, obtained with GALFIT. For further information about the fit results, see Table \ref{Allisons}. Right panel: Residuals from the Sersic fit. We recognize a bright core, which is not aligned with the disk and leaves a residual. Furthermore, there is another additional component in the northern part of the galaxy. In the lower panels we show the multi-component fit described in section \ref{GALFIT_Section} which includes those extra components. \label{JR_fig_M}}
\end{figure}

Of the two targets in this study, MACS2129-1 shows a clearly extended morphology, which is most prominent on the HST F160W image. We use this image to reconstruct the shape of the object in the source plane, based on the determined redshift. We show the resulting de-lensed image in Fig. \ref{JR_fig_M}. Similarly, we use the HST F850LP image of A1413 to reconstruct the unlensed image of A1413-1, which is shown in Fig. \ref{JR_fig_A}.

We use GALFIT version 3.0 \citep{2002AJ....124..266P,2010AAS...21522909P} to fit a 2D Sersic model convolved with the reconstructed PSF, to the de-lensed F160W image of MACS2129-1, and to the F850LP image of A1413-1. The code is run on 100 Monte Carlo realizations of the reconstructions. In the case of MACS2129-1, three of the 100 realizations give highly deviant results with very high $\chi^{2}$, which we subsequently discard. Uncertainties are computed as the standard deviation of all (remaining) results. The best-fit parameters (effective half-light radius $r_{eff}$, Sersic index $n$, and axis ratio $\frac{b}{a}$) are then used to quantify the structure of the galaxies. We compute the circularized radii as $r_{c}=r_{eff}\cdot\sqrt{\frac{b}{a}}$. The Sersic index provides an indication whether the light profile resembles more an exponential disk profile ($n=1$) or an elliptical galaxy ($n=4$). The results are summarized in Table \ref{Allisons} and visualized in Fig. \ref{JR_fig_A} and Fig \ref{JR_fig_M}.
As can be seen in the upper half of Fig. \ref{JR_fig_M}, the one-component Sersic fit to MACS2129-1 leaves two residuals, one in the core and one in the northern part of the galaxy. To account for these deviations from a simple Sersic profile, we produced a second fit, which models the two extra components with a psf, i.e. a point-source-like component. The lower half of Fig. \ref{JR_fig_M} shows the residuals of this better fit. We note that the multi-component fit does not significantly alter the result for the Sersic index and effective radius of MACS2129-1.

We use the definition of \cite{2012arXiv1206.5000B} of compactness (at $z\sim2$)

\begin{equation}
\log~M_\star[M_{\odot}]\cdot{R[kpc]^{-1.5}}>10.3
\end{equation}

\noindent to assess how compact the two galaxies in our study are. Here R corresponds to the circularized effective radius. It is thus confirmed that MACS2129-1 is indeed a compact galaxy according to that definition, with $\log~M_\star[M_{\odot}]\cdot{r_{c}[kpc]^{-1.5}}=10.67\pm0.28$.
The compactness of A1413-1 is computed as $\log~M_\star[M_{\odot}]\cdot{r_{c}[kpc]^{-1.5}}=10.54\pm0.14$ and thus also fulfills the criterion for compactness.

\begin{deluxetable}{@{}ccccccc@{}}
\tablecaption{GALFIT fitting results for the reconstructed images of the lensed galaxies A1413-1 and MACS2129-1 \label{Allisons}}
\tablewidth{0pt}
\tablehead{
Parameter & A1413-1 & MACS2129-1
}

\startdata

$n$ & $3.25\pm0.02$ & $0.88\pm0.01$ \\
$r_{e}$ & $2.38\pm0.01kpc$ & $2.60\pm0.29kpc$ \\
$\frac{b}{a}$ & $0.50\pm0.01$ & $0.47\pm0.03$ \\
$r_{c}$ & $1.68\pm0.01kpc$ & $1.78\pm0.14kpc$ \\
PA & $14^{\circ} EofN$ & $-34^{\circ} EofN$ \\
\hline
\enddata
\end{deluxetable}


\section{Discussion and conclusions}
\label{Discussion_Section}

In this article we analyzed the spectra of two gravitationally-lensed distant compact quiescent galaxies and derived their redshifts, stellar masses and ages, as well as constraints on dust extinction and potential star formation, from SED fitting. We also analyzed their morphologies based on HST images which we de-lensed with available mass-models for the lensing clusters. Interpretations of the results are presented below.

\subsection{Redshifts and derivation of intrinsic magnitudes}
\label{Magnifications_Section}

\begin{figure}[h!]
\includegraphics[angle=0,scale=1.0,width=1.0\linewidth]{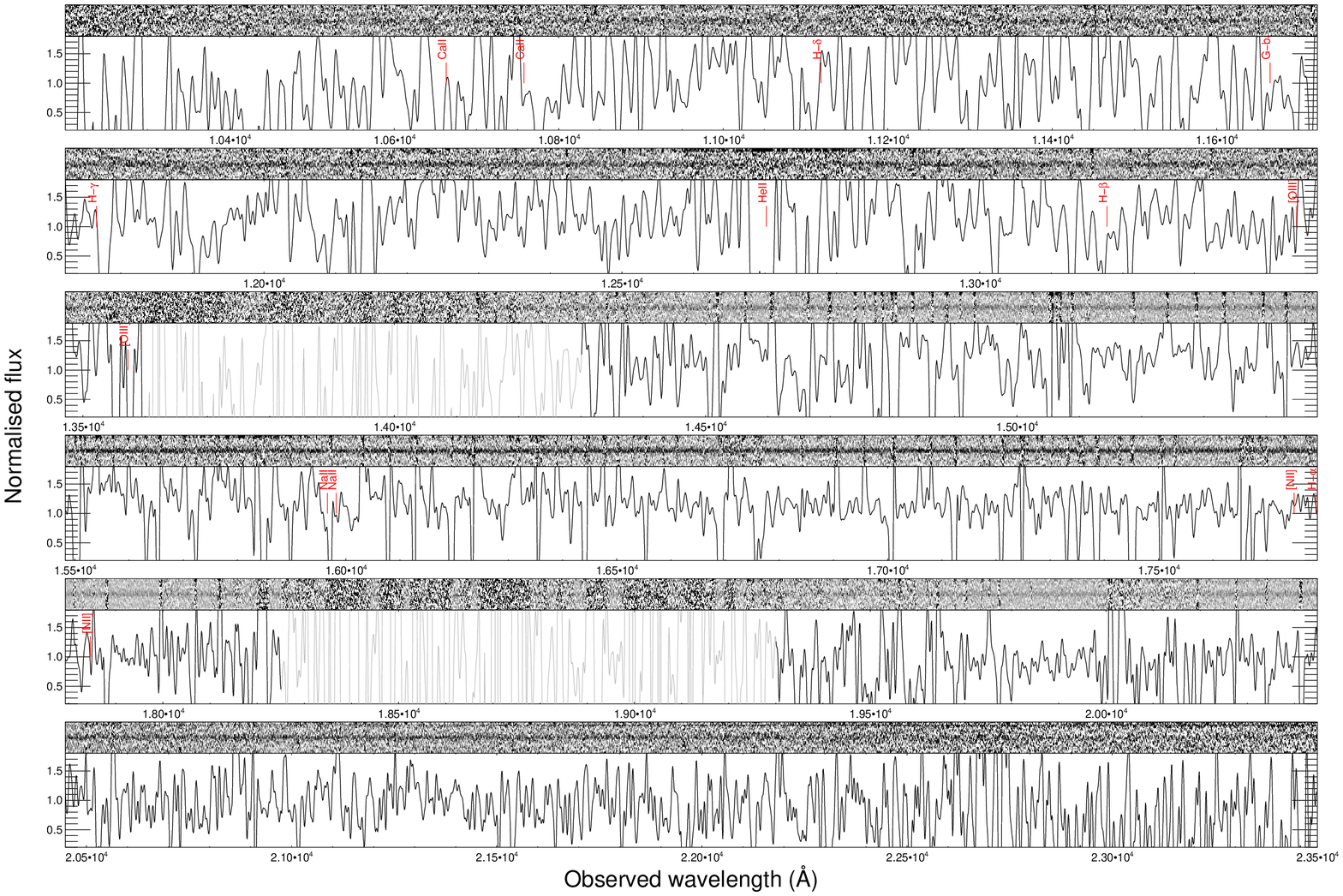}
\caption{The 2D and 1D X-Shooter NIR spectrum of A1413-1. We overplot the expected positions of several spectral lines at the redshift of $z=1.71$. Little black lines above the 1D spectrum indicate skylines. The wavelength regions in the gaps between the J- and H-bands and the H- and K-bands are plotted in grey to avoid distraction by the high noise (and absence of information) there. \label{spec_a1413}}
\end{figure}

\begin{figure}[t!]
\includegraphics[angle=0,scale=1.0,width=1.0\linewidth]{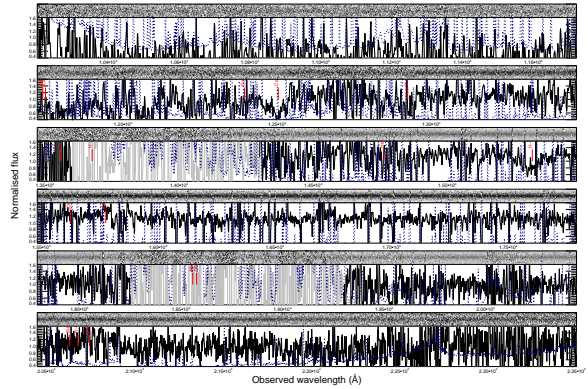}
\caption{The 2D and 1D X-Shooter NIR spectrum of MACS2129-1. We overplot the expected positions of several spectral lines at the redshift of $z=2.15$. The blue line indicates the error spectrum. Little black lines above the 1D spectrum indicate skylines. The wavelength regions in the gaps between the J- and H-bands and the H- and K-bands are plotted in grey to avoid distraction by the high noise (and absence of information) there. \label{spec_macs2129}}
\end{figure}

In the SED fitting procedure, the Balmer/4000\AA-break is the feature which enables a reliable and robust determination of the redshifts of galaxies with evolved stellar populations. The resulting redshifts are $z=1.71\pm0.02$ for A1413-1 and $z=2.15\pm0.01$ for MACS2129-1. The respective error bars are taken from the $68\%$ confidence intervals of the best fits. To visualize where spectral lines are expected to be at these redshifts, we overplot their positions at those redshifts in Figures \ref{spec_a1413}, \ref{spec_macs2129}, and \ref{specs_halpha}. We also use the derived redshifts to determine the magnifications according to the available mass models for both clusters. The resulting lensing magnifications are $1.73\pm0.05$ for A1413-1 and $3.95\pm0.95$ for MACS2129-1 and thus lower than the initial estimations based on the assumption of $z\sim2$. Adopting $K\star\sim21.4$ as the typical luminosity ($L\star$) of red galaxies at $z\sim2$ \citep{2007ApJ...656...42M}, the two galaxies in this study exhibit luminosities of $\sim5.2\times L\star$ (A1413-1) and $\sim6.3\times L\star$ (MACS2129-1). Their stellar masses correspond to $\sim0.9\times~M\star$ and $\sim1.3\times~M\star$ \citep{2009ApJ...701.1765M}. The de-lensed magnitudes according to the lensing magnifications were used to scale the stellar masses and star formation rates of the SED fits.

\subsection{Constraints on line-fluxes and star formation}
\label{LineFlux_Section}

\begin{figure}[t!]
\includegraphics[angle=0,scale=0.4,width=1.0\linewidth]{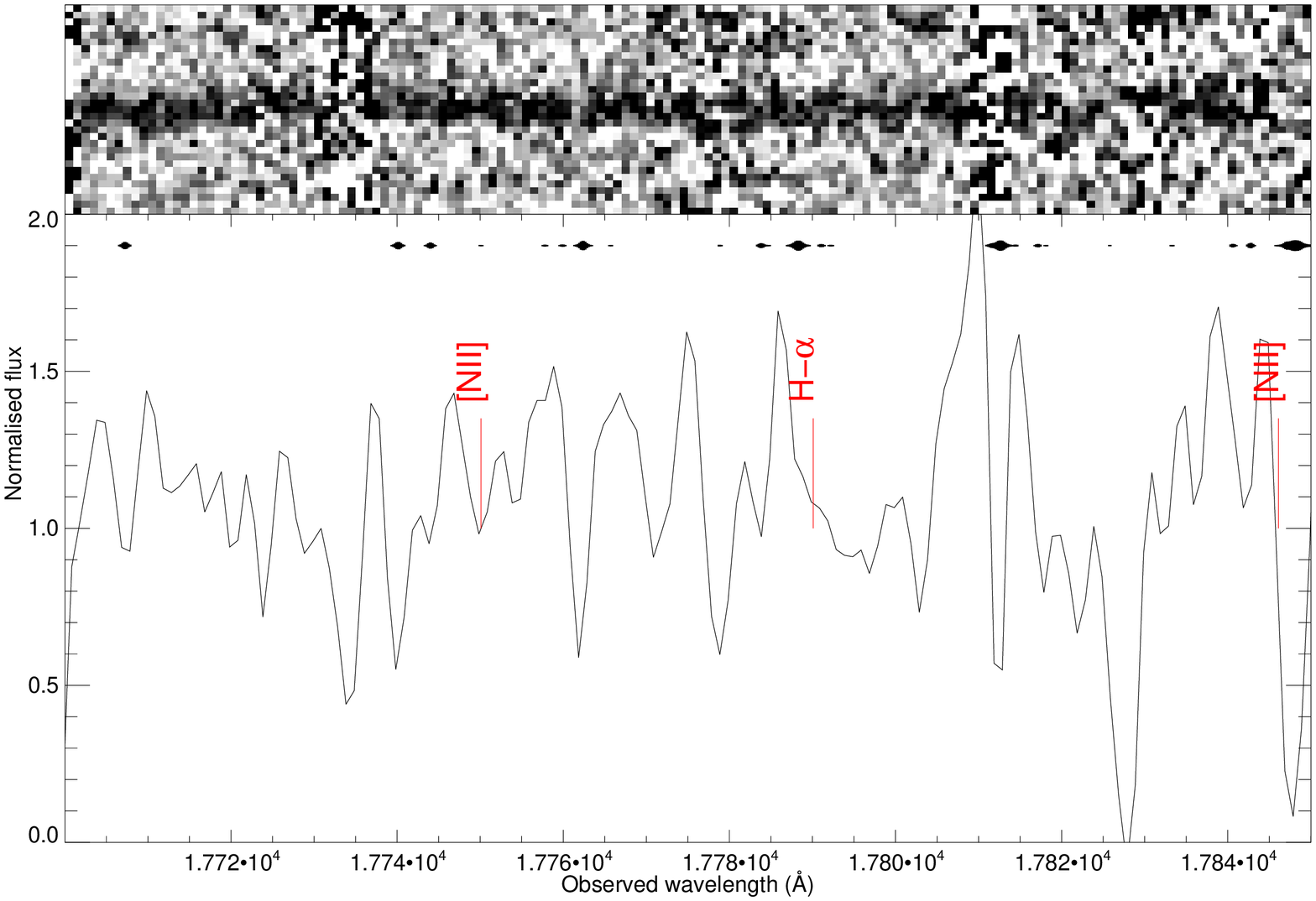}
\includegraphics[angle=0,scale=0.4,width=1.0\linewidth]{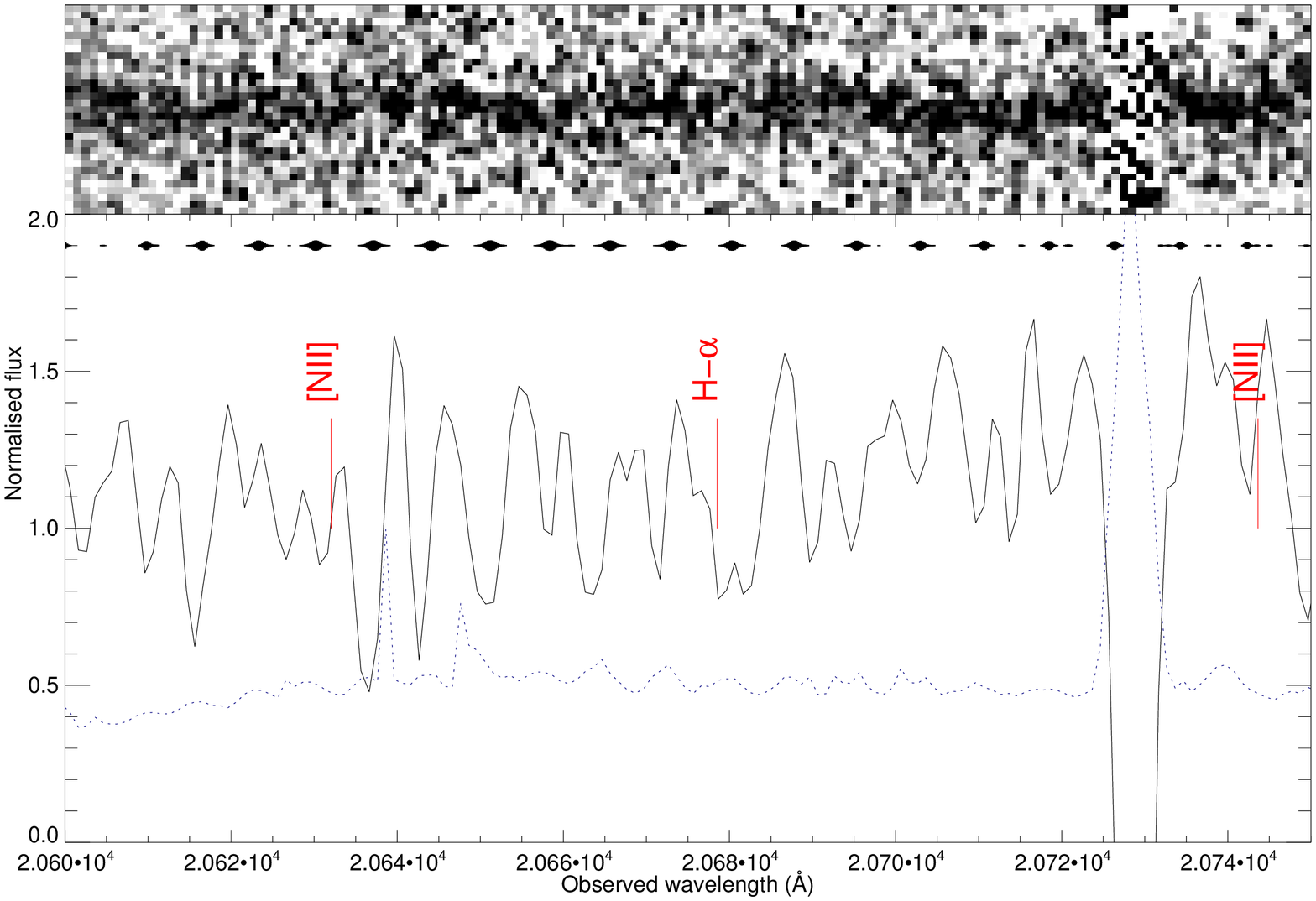}
\caption{Upper panel: Zoom-in on the wavelength region around the expected position of the $H\alpha$ line in the 2D and 1D spectrum of A1413-1. Lower panel: Zoom-in on the wavelength region around the expected position of the $H\alpha$ line in the 2D and 1D spectrum of MACS2129-1. The cutouts show that there is no detectable emission line. \label{specs_halpha}}
\end{figure}

In Fig. \ref{specs_halpha} we demonstrate the non-detectability of potential $H\alpha$ emission by zooming in on the wavelength regions where $H\alpha$ is expected at the redshifts of the two galaxies. In both the 1D and 2D spectral cutouts it is clearly visible that there is no detectable emission line.
In order to constrain the amount of star formation which might have been still on-going in the two galaxies, despite the lack of detectable emission lines, we determined upper limits on potential $H\alpha$ line emission in the spectra. For that purpose, we subtracted the continuum around the positions where we would expect the $H\alpha$ line at the respective redshifts and added artificial emission lines. In this way, we infer $3\sigma$ limits on $H\alpha$ line emission of $2\cdot10^{17}erg\cdot s^{-1}\cdot cm^{-2}$ and $2.4\cdot10^{17}erg\cdot s^{-1}\cdot cm^{-2}$ for A1413-1 and MACS2129-1, respectively. Converting into intrinsic potential $H\alpha$ luminosities (taking also the lensing magnification into account) and applying the \cite{1998ARA&A..36..189K} relation, the $3\sigma$ upper limits on star formation rates in A1413-1 and MACS2129-1 turn out to be $<1.8\cdot M_{\odot}\cdot yr^{-1}$ and $<1.7\cdot M_{\odot}\cdot yr^{-1}$, respectively.
The Kennicutt relation is based on the assumption of a \cite{1955ApJ...121..161S} IMF and solar metallicities. Converting to the assumed Chabrier IMF via division by a factor of 1.58 \citep{2007ApJS..173..256T}, the $3\sigma$ upper limits are of the order of $1 M_{\odot}\cdot yr^{-1}$. These are the best limits on potentially on-going star-formation in quiescent $z\sim2$ galaxies so far, even compared to the most recent similar studies \citep{2009ApJ...700..221K,2010ApJ...715L...6O,2011ApJ...736L...9V,2012ApJ...754....3T}.


\subsection{Stellar population properties}
\label{Stellar_Section}

From SED modeling, we derived the ages of the stellar populations to be in the range of $2.4^{+0.8}_{-0.6}$~Gyr (A1413-1) and $1.7\pm0.3$~Gyr (MACS2129-1).
This is an intriguing result, as they turn out to be older and thus exhibit higher mass-to-light ratios than most galaxies from other studies \citep{2008ApJ...677..219K,2009ApJ...700..221K,2010ApJ...718L..73V,2011ApJ...736L...9V,2012ApJ...754....3T,2012arXiv1211.3424V}. This is mostly due to the fact that younger galaxies are easier to detect. \cite{2012arXiv1211.3424V} indeed argue that their sample is biased towards young galaxies, compared to a mass-limited sample.
By scaling to the intrinsic luminosities, taking into account the enhancement of the observed brightness by the lensing effect, and including the uncertainty on the lensing magnification in quadrature into the error budget, we infer their stellar masses to be $(7.6\pm1.1)\cdot10^{10}~M_{\odot}$ (A1413-1) and $(1.1\pm0.28)\cdot10^{11}~M_{\odot}$ (MACS2129-1). This makes them two of the least (intrinsically) luminous  quiescent high-z galaxies whose properties have been studied spectroscopically. As visualized in Fig. \ref{mass_size_plot} this mass range has been probed before, which was however only possible with exorbitant use of telescope time ($\sim500h$ for the GMASS sample). The lensing approach makes these studies much more feasible, and will in the future also give us the possibility to test the properties of larger samples of $z\sim2$ quiescent $~M\star$ (and even $~L\star$) galaxies, which are much more representative of the overall population than the brightest ones (see also \cite{2012arXiv1211.3424V} for a more detailed discussion of this issue). 

\subsection{Morphologies}
\label{Morphology_Section}

In section~\ref{GALFIT_Section} we described the fitting of surface brightness profiles with the GALFIT tool, the results of which we summarize in Table \ref{Allisons}. For A1413-1 we find a relatively high Sersic index of $n\sim3.3$ which resembles an early-type galaxy. The structural analysis of galaxy MACS2129-1 reveals a profile which is more disky than an exponential disk with $n=0.88$. This shape classifies it as an S0 like galaxy. Only relatively recently it has been possible to identify disk-like galaxies around $z\sim2$ \citep[e.g.][]{2004ApJ...605...37S,2008ApJ...672..146S,2006Natur.442..786G, 2011ApJ...730...38V, 2013ApJ...762...83C}. Combined with the constraints on potentially on-going star formation it constitutes the first spectroscopically confirmed passive lenticular galaxy at $z>2$.





\subsection{Stellar masses and sizes}
\label{Mass_Size_Section}

\begin{figure}[t!]
\includegraphics[angle=90,scale=1.0,width=1.0\linewidth]{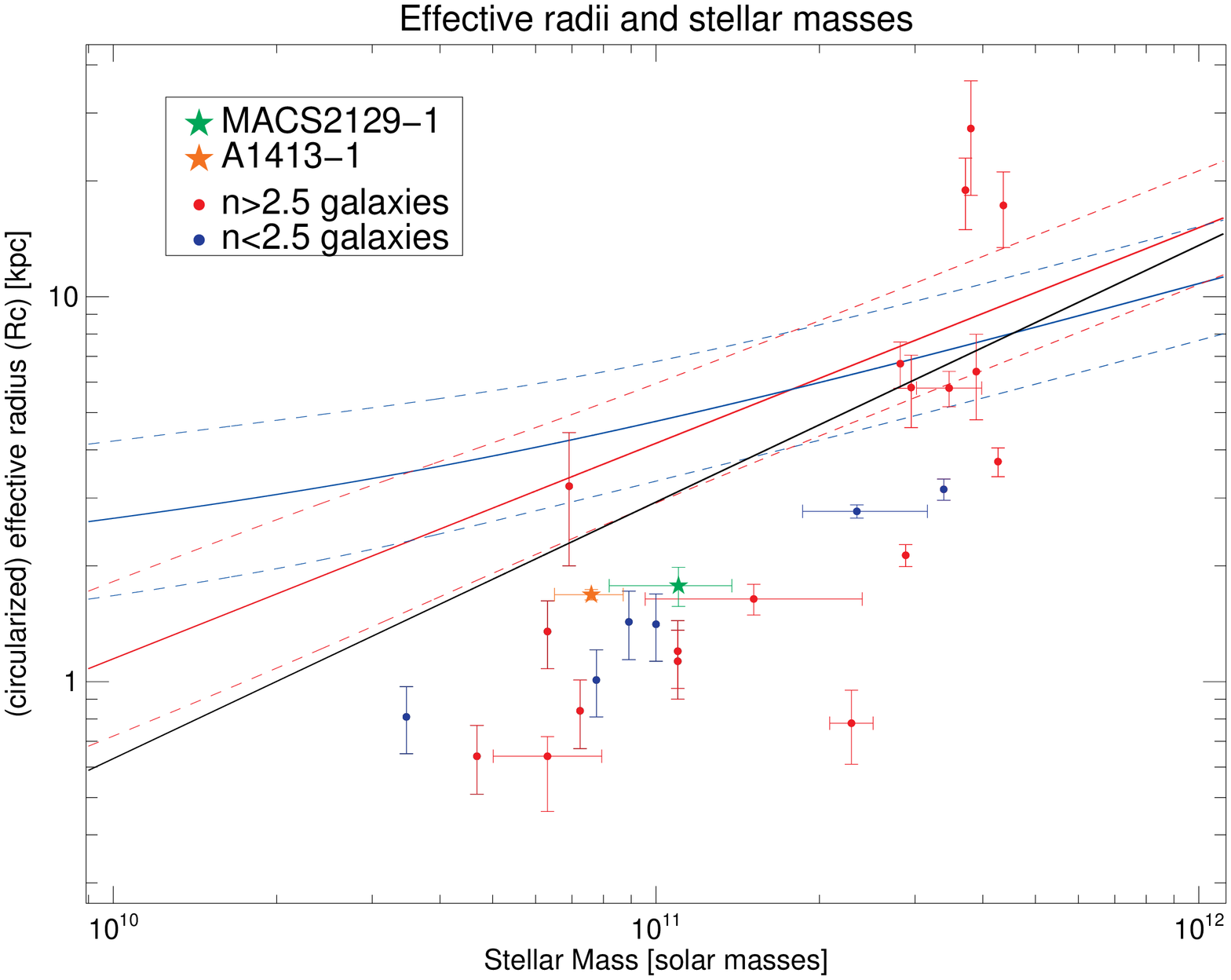}
\caption{Effective (circularized) radii of the lensed $z\sim2$ galaxies A1413-1 and MACS2129-1 plotted against their stellar masses. For comparison we overplot the values for a sample of 24 $1.5<z<2.5$ passive galaxies compiled from \citet{2012ApJ...754....3T}, \citet{2011ApJ...736L...9V}, \citet{2009ApJ...700..221K,2009Natur.460..717V}, \citet{2010ApJ...718L..73V}, \citet{2010ApJ...715L...6O}, \citet{2012arXiv1211.0280M}, \citet{2010MNRAS.401..933M}, and \citet{2008A&A...482...21C}. The local stellar mass-size-relations for early-type (red color) and late-type (blue color) galaxies from \citet{2003MNRAS.343..978S} are also overplotted, with their intrinsic scatter in dashed lines. The compactness criterion of \cite{2012arXiv1206.5000B} is indicated by the bold black line.}
\label{mass_size_plot}.
\end{figure}

In order to compare the stellar masses and sizes of the two galaxies in our study, we compile a sample of passive galaxies in the redshift range $1.5<z<2.5$ from the literature. 
All these studies used the Chabrier IMF \citep{2003PASP..115..763C} and thus provide some useful comparison of stellar masses, and estimate their (circularized) half-light radii via Sersic profile fits. The 24 galaxies in this comparison sample are taken from \citet{2012ApJ...754....3T}, \citet{2011ApJ...736L...9V}, \citet{2009ApJ...700..221K,2009Natur.460..717V}, \citet{2010ApJ...718L..73V}, \citet{2010ApJ...715L...6O}, \citet{2012arXiv1211.0280M}, \citet{2010MNRAS.401..933M}, and \citet{2008A&A...482...21C}.
For comparison with the typical properties of galaxies in today's Universe, in Fig. \ref{mass_size_plot} we plot the local stellar mass-size-relations for early-type (red color) and late-type (blue color) galaxies from \citet{2003MNRAS.343..978S}, with their intrinsic scatter plotted in dashed lines. The same color scheme is used to divide the $1.5<z<2.5$ galaxy sample into early-type ($n>2.5$) and late-type ($n<2.5$) galaxies. The two galaxies from our study are plotted in green (MACS2129-1) and orange (A1413-1) colors.
As found in previous studies, quiescent galaxies at $1.5<z<2.5$ are significantly more compact than galaxies in today's local Universe. The compactness criterion of \cite{2012arXiv1206.5000B} is indicated by the bold black line.

The intriguing diversity of quiescent galaxies at $z\sim2$ \citep[see also][]{2010ApJ...715L...6O} clearly indicates that the spectroscopic sample of quiescent $z\sim2$ galaxies needs to be extended significantly to establish a robust picture of their role in the formation and evolution of early-type galaxies at all redshifts.

\subsection{Prospects for future detailed studies}
\label{Prospects_Section}

An additional advantage of observing lensed high-redshift galaxies is the increased spatial
resolution due to the gravitational lensing effect. For unlensed compact $z\sim2$ galaxies there is
essentially no spatial information available in ground based observations,
given the extreme compactness of these objects, \citep[e.g.][]{2012ApJ...754....3T}. 
The additional spatial information is in particular striking for MACS2129-1
that is stretched along a caustic line resulting in an observed extension of almost
$2\arcsec$. With a deeper observation using multiple slit positions it should be possible to determine both the dependence of the (line-of-sight) velocity dispersion
on position within the galaxy as well as to establish if there is rotation of the
overall system \citep{2013arXiv1305.0268B}. We point out that the lensing approach will greatly increase the feasibility of continuum and absorption line spectroscopy of faint, i.e. $\sim L_{\star}$ quiescent galaxies at high redshifts ($z\gtrsim2$).

We emphasize the interesting finding that MACS2129-1 exhibits a surface
brightness profile which strongly resembles a disk galaxy, which corroborates
the existence of  a significant fraction of massive passive disks at $z\sim2$.
The question whether disk-like galaxies dominate the passive galaxy population
at $1.5<z<2.5$ is still a highly-debated topic, with no firm conclusion at hand
yet \citep[e.g.][]{2011ApJ...730...38V,2011ApJ...742...96W,2013MNRAS.428.1460B}.

In this current study, we have demonstrated that even in cases with limited S/N
it is possible to use the
continuum emission to derive useful constraints on important properties like
redshifts, stellar masses and ages as well as the resulting formation
redshifts. Subsequent studies of
larger samples of quiescent $z\sim2$ galaxies with intrinsic luminosities down
to $L\star$ will allow for statistically more robust insights into
relationships between (stellar) masses, sizes, morphologies, and formation
redshifts. \\

\acknowledgments

We thank Teddy Frederiksen, Martin Sparre, Antonio de Ugarte Postigo, and Stefano Zibetti for helpful
discussions. We thank the anonymous referee for constructive and helpful reviews. SG thanks Gottfried Beyvers for detailed proof-reading of the manuscript. We gratefully acknowledge support from the Lundbeck foundation, and
the Dark Cosmology Centre which is funded by the Danish National Research
Foundation. JR is supported by the EU Career Integration Grant 294074. TK acknowledges support by the European Commission under the Marie Curie Intra-European Fellowship Programme. JPUF acknowledges support from the ERC-StG grant EGGS-278202.



{\it Facilities:} \facility{VLT:Kueyen}, \facility{VLT:Antu}, \facility{HST (ACS, WFC3)}, \facility{Hale}.




\bibliographystyle{apj}
\bibliography{bibtex_refs31}{}

\appendix

\begin{deluxetable}{@{}ccccccc@{}}
\tablecaption{Spectro-photometric measurements for A1413-1. $^{(a)}$
Effective wavelength calculated as in Eq. 5.  $^{(b)}$ Short and long
wavelength ends of the spectro-photometric bins. $^{(c)}$ Observed AB
magnitude. $^{(d)}$ Intrinsic AB magnitude inferred from the lensing
magnification. \label{tab_a1413-1}}
\tablewidth{0pt}
\tablehead{
Facility & Filter/$\lambda_{eff}[nm]^{(a)}$ & $cut-on^{(b)}$ & $cut-off^{(b)}$ & $mag_{obs}^{(c)}$ & $\Delta mag$ & $mag_{int}^{(d)}$ 
}

\startdata

\hline
HST/ACS & F775W  & - & - & 24.25 & 0.05 & 24.85 \\ 
HST/ACS & F850LP & - & - & 23.25 & 0.05 & 23.85 \\
XSH/NIR & 1044.09 & 1025.0 & 1055.0 & 22.634 & 0.216 & 23.234  \\
XSH/NIR & 1068.59 & 1055.0 & 1085.0 & 22.306 & 0.147 & 22.906 \\
XSH/NIR & 1098.27 & 1085.0 & 1113.0 & 21.992 & 0.103 & 22.592 \\
XSH/NIR & 1164.83 & 1136.0 & 1190.0 & 21.885 & 0.073 & 22.485 \\
XSH/NIR & 1207.11 & 1190.0 & 1225.0 & 21.735 & 0.073 & 22.335 \\
XSH/NIR & 1238.78 & 1225.0 & 1250.0 & 21.651 & 0.068 & 22.251 \\
WIRC & J & - & - & 21.93 & 0.12 & 22.53  \\
XSH/NIR & 1259.32 & 1250.0 & 1275.0 & 21.611 & 0.087 & 22.211 \\
XSH/NIR & 1286.26 & 1275.0 & 1295.0 & 21.485 & 0.091 & 22.085 \\
XSH/NIR & 1305.40 & 1295.0 & 1315.0 & 21.548 & 0.069 & 22.148 \\
XSH/NIR & 1324.02 & 1315.0 & 1335.0 & 21.612 & 0.079 & 22.212 \\
XSH/NIR & 1456.64 & 1440.0 & 1474.3 & 21.347 & 0.056 & 21.947 \\
XSH/NIR & 1488.54 & 1474.4 & 1498.5 & 21.185 & 0.056 & 21.785 \\
XSH/NIR & 1511.31 & 1498.6 & 1521.0 & 21.185 & 0.059 & 21.785 \\
XSH/NIR & 1541.88 & 1521.0 & 1562.2 & 21.084 & 0.031 & 21.684 \\
XSH/NIR & 1587.96 & 1562.3 & 1609.7 & 21.079 & 0.031 & 21.679 \\
XSH/NIR & 1627.22 & 1609.8 & 1645.0 & 21.001 & 0.031 & 21.601 \\
XSH/NIR & 1661.47 & 1645.1 & 1679.3 & 21.038 & 0.031 & 21.638 \\
XSH/NIR & 1692.84 & 1679.4 & 1708.6 & 21.011 & 0.031 & 21.611 \\
XSH/NIR & 1722.86 & 1708.7 & 1737.1 & 20.922 & 0.031 & 21.522 \\
XSH/NIR & 1751.04 & 1737.2 & 1763.7 & 20.964 & 0.031 & 21.564 \\
XSH/NIR & 1781.63 & 1763.8 & 1798.0 & 20.955 & 0.035 & 21.555 \\
XSH/NIR & 1984.47 & 1971.0 & 1995.0 & 20.928 & 0.052 & 21.528 \\
XSH/NIR & 2035.02 & 2021.0 & 2048.0 & 20.733 & 0.046 & 21.333 \\
XSH/NIR & 2081.69 & 2058.0 & 2100.0 & 20.710 & 0.050 & 21.310 \\
XSH/NIR & 2130.80 & 2112.0 & 2149.0 & 20.656 & 0.048 & 21.256 \\
WIRC & Ks (2147.40) & - & - & 20.839 & 0.064 & 21.439  \\
XSH/NIR & 2179.26 & 2151.0 & 2211.0 & 20.692 & 0.050 & 21.292 \\
XSH/NIR & 2232.73 & 2212.0 & 2260.0 & 20.693 & 0.078 & 21.293 \\
XSH/NIR & 2287.43 & 2275.0 & 2300.0 & 21.006 & 0.186 & 21.606 \\
\hline
\enddata
\end{deluxetable}

\begin{deluxetable}{@{}ccccccc@{}}
\tablecaption{Spectro-photometric measurements for MACS2129-1. $^{(a)}$
Effective wavelength calculated as in Eq. 5.  $^{(b)}$ Short and long
wavelength ends of the spectro-photometric bins. $^{(c)}$ Observed AB
magnitude. $^{(d)}$ Intrinsic AB magnitude inferred from the lensing
magnification. \label{tab_macs2129-1}}
\tablewidth{0pt}
\tablehead{
Facility & Filter/$\lambda_{eff}[nm]^{(a)}$ & $cut-on^{(b)}$ & $cut-off^{(b)}$ & $mag_{obs}^{(c)}$ & $\Delta mag$ & $mag_{int}^{(d)}$ 
}

\startdata

\hline
HST/ACS & F606W  & - & - & 23.7 & 0.3 & 25.21 \\
HST/WFC3 & F105W & - & - & 22.14 & 0.15 & 23.65 \\
XSH/NIR & 1064.01 & 1025.0 & 1095.0 & 22.108 & 0.087 & 23.618 \\
HST/WFC3 & F110W & - & - & 21.36 & 0.10 & 22.87 \\
XSH/NIR & 1130.41 & 1095.0 & 1165.0 & 21.889 & 0.078 & 23.399 \\ 
XSH/NIR & 1190.63 & 1165.0 & 1215.0 & 21.319 & 0.046 & 22.829 \\ 
HST/WFC3 & F125W & - & - & 20.90 & 0.09 & 22.41 \\
XSH/NIR & 1228.36 & 1215.0 & 1240.0 & 20.977 & 0.045 & 22.487 \\
XSH/NIR & 1249.32 & 1240.0 & 1260.0 & 20.973 & 0.044 & 22.483 \\ 
ISAAC & J (1250.5) & - & - & 20.914 & & 22.424 \\ 
XSH/NIR & 1276.50 & 1260.0 & 1290.0 & 20.709 & 0.043 & 22.219 \\ 
XSH/NIR & 1298.29 & 1290.0 & 1305.0 & 20.727 & 0.044 & 22.237 \\
XSH/NIR & 1312.77 & 1305.0 & 1320.0 & 20.596 & 0.034 & 22.106 \\ 
XSH/NIR & 1327.19 & 1320.0 & 1335.0 & 20.593 & 0.041 & 22.103 \\
HST/WFC3 & F140W & - & - & 20.33 & 0.06 & 21.84 \\ 
XSH/NIR & 1457.87 & 1440.0 & 1476.3 & 20.328 & 0.022 & 21.838 \\ 
XSH/NIR & 1492.49 & 1476.4 & 1504.4 & 20.262 & 0.022 & 21.772 \\ 
XSH/NIR & 1514.68 & 1504.5 & 1521.0 & 20.245 & 0.028 & 21.755 \\ 
XSH/NIR & 1544.03 & 1521.0 & 1566.8 & 20.259 & 0.015 & 21.769 \\ 
XSH/NIR & 1588.99 & 1566.9 & 1608.2 & 20.115 & 0.015 & 21.625 \\ 
HST/WFC3 & F160W & - & - & 20.16 & 0.07 & 21.626 \\ 
XSH/NIR & 1626.76 & 1608.3 & 1646.4 & 20.170 & 0.015 & 21.680 \\
XSH/NIR & 1660.86 & 1646.5 & 1677.7 & 20.082 &  0.015 & 21.592 \\ 
XSH/NIR & 1688.87 & 1677.8 & 1702.2 & 20.059 & 0.015 & 21.569 \\ 
XSH/NIR & 1717.68 & 1702.3 & 1733.6 & 20.041 & 0.015 & 21.551 \\ 
XSH/NIR & 1746.74 & 1733.7 & 1758.4 & 19.965 &  0.015 & 21.475 \\ 
XSH/NIR & 1775.39 & 1758.5 & 1794.4 & 19.973 & 0.015 & 21.483 \\ 
XSH/NIR & 1984.37 & 1971.0 & 1995.0 & 19.860 & 0.023 & 21.370 \\ 
XSH/NIR & 2034.97 & 2021.0 & 2048.0 & 19.779 &  0.022 & 21.289 \\ 
XSH/NIR & 2081.61 & 2058.0 & 2100.0 & 19.658 & 0.022 & 21.168 \\ 
XSH/NIR & 2130.72 & 2112.0 & 2149.0 & 19.708 & 0.024 & 21.218 \\ 
ISAAC & Ks (2165.2) & - & - & 19.704 & 0.01 & 21.214 \\ 
XSH/NIR & 2179.10 & 2151.0 & 2211.0 & 19.712 & 0.024 & 21.222 \\ 
XSH/NIR & 2232.38 & 2212.0 & 2260.0 & 19.751 & 0.039 & 21.261 \\ 
XSH/NIR & 2287.42 & 2275.0 & 2300.0 & 19.443 & 0.049 & 20.953 \\ 
\hline
\enddata
\end{deluxetable}

\end{document}